\documentclass{article}
\usepackage{spconf}

\usepackage{stmaryrd}

\usepackage{float}

\usepackage[english]{babel} 
\usepackage[utf8]{inputenc}
\usepackage[]{graphicx}
\usepackage{caption}
\usepackage{subcaption}

\usepackage{mathtools}
\usepackage[T1]{fontenc}
\usepackage{amsmath, amsthm, amssymb}

\usepackage[ruled]{algorithm}
\usepackage{algorithmic}

\title{Schedule based Self Localization of asynchronous wireless nodes with experimental validation  }
\name{Baptiste Cavarec, Satyam Dwivedi,  Mats Bengtsson, and Peter H\"andel}
\address{Department of Signal Processing, School of Electrical Engineering \\KTH Royal Institute of Technology, Stockholm, Sweden }
\date{September 2016}
  
\begin{document}

\maketitle
\begin{abstract}
        In this paper we have proposed clock error mitigation from the measurements in the scheduled based self localization system. We propose measurement model with clock errors while following a scheduled transmission among anchor nodes. Further, RLS algorithm is proposed to estimate clock error and to calibrate measurements of self localizing node against relative clock errors of anchor nodes. A full-scale experimental validation is provided based on commercial off-the-shelf UWB radios under IEEE-standardized protocols.
\end{abstract}
\keywords{self-localization, indoor positioning, UWB}
\section{Introduction}
In the last decade, the interest for positioning systems have skyrocketed, both in commercial and industrial environments. In this field, Ultra-Wide Band devices seems promising since it has potential for decimeter precision \cite{4653424,6471216}.   

 
 Schedule based positioning has been proposed and analysed in \cite{6340491, 6510560, 6490337, Zachariah2014}, as a method allowing asynchronous nodes in a network to perform cooperative localization without communication overhead.
In scheduled based positioning, every node takes turn to broadcast its messages as per a schedule. The schedule is a-priori decided. In \cite{6340491} and \cite{6510560}, schedule based positioning is used for distributed position estimation when every node can estimate position of every other node in the network including itself. In \cite{6490337} and \cite{Zachariah2014}, schedule based positioning is used for self localization of a node while only receiving scheduled transmissions from anchor nodes. In both the contexts of distributed positioning and self localization, anchor positions are assumed to be known. Moreover, the system can be implemented with low-cost devices. In addition, no wired or wireless synchronization is needed.

 The idea of schedule based positioning relies on generating analog delays $\delta$ at every node. Whereas, in practice these delays are generated using digital discrete clocks. The clocks also suffer nonidealities such as clock skew and clock jitter \cite{DwivediEtAl2015_jointsyncrange}. In this paper we have suggested and demonstrated implementation of schedule based self localization scheme using commercial wireless ranging devices. The device used is the Decawave devices \cite{decref, 7492183},working under IEEE standards.Thus, easy adoption of the technology is provided. 
 We have proposed practical measurement model for schedule based measurements.  Thereupon we have proposed calibration scheme to minimise clock errors from the measurements. Time difference of arrival (TDOA) measurements at self localizing node in an experimental setup is collected. 
 To the authors knowledge, there are no other experimental validation available of passive scalable and asynchronous ultra wide band localization system.
 The collected measurements are  calibrated to minimise clock errors. The calibrated measurements are then used to infer position of the self localizing node. Further, performance of position estimation using measurements is compared with simulated performance and Cram\'er-Rao lower bound on position estimation.   

\section{Measurements model}
\label{sec:model}
\subsection{Two Way Ranging(TWR) and Passive Listening}

\begin{figure}[h!]
        \centering
        \includegraphics[width=0.5\columnwidth]{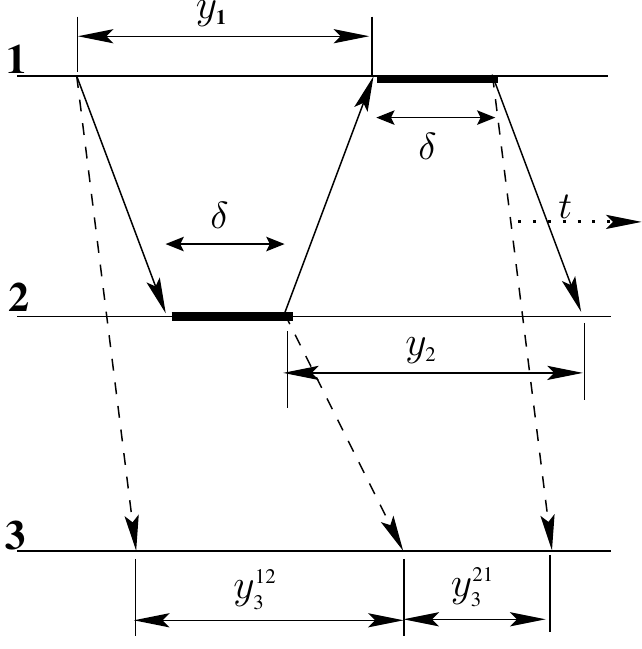}
        \caption{Node \textbf{1}, node \textbf{2} and node \textbf{3} two way ranging diagram}
        \label{fig:SDSTWRList}
    \end{figure}

Figure \ref{fig:SDSTWRList} is the timing diagram of node \textbf{1} and node  \textbf{2} participating in a two-way ranging measurement while  node \textbf{3} is doing TDOA measurements of the signal it receives from node \textbf{1} and node \textbf{2}.  As shown in the figure, node \textbf{1} sends a \textsc{Ping} and node \textbf{2} replies with a \textsc{Respond} after a delay of $\delta_{\text{\tiny B}}$. Again node \textbf{1} transmits after a delay $\delta_{\text{\tiny A}}$. The distance between node \textbf{1} and node \textbf{2} is $\rho_{\text{\tiny \textbf{12}}}$. node \textbf{1} and node \textbf{2} measure the round trip times (RTT) $y_{\text{\tiny \textbf{1}}}$, and $y_{\text{\tiny \textbf{2}}}$. Node \textbf{3} in the figure is the passive listening node and does measurements $y_{\text{\tiny \textbf{3}}}^{\text{\tiny \textbf{12}}}$ and $y_{\text{\tiny \textbf{3}}}^{\text{\tiny \textbf{21}}}$. 
Node \textbf{1} and node \textbf{2} introduce delay $\delta$, which is a known parameter in the network,  by counting number of periods of their respective clocks $\mathcal{C}_{\text{\tiny \textbf{1}}}$ and $\mathcal{C}_{\text{\tiny \textbf{2}}}$. Clocks at node \textbf{1} and node \textbf{2} are modelled as $\mathcal{C}_{\text{\tiny \textbf{1}}} = (1 + \vartheta_{\text{\tiny \textbf{1}}}) t$ and $\mathcal{C}_{\text{\tiny \textbf{2}}} = (1 + \vartheta_{\text{\tiny \textbf{2}}}) t$. Where $\vartheta_{\text{\tiny \textbf{1}}}$ and $\vartheta_{\text{\tiny \textbf{2}}}$ are respective clock skews. $t$ is the absolute reference time. The delays at node \textbf{1} and node \textbf{2} are often imprecise due to hardware constraints. Thus, the actual delays are given as $\Delta_{\text{\tiny \textbf{1}}} = \delta + \epsilon_{\text{\tiny \textbf{1}}}$ and $\Delta_{\text{\tiny \textbf{2}}} = \delta + \epsilon_{\text{\tiny \textbf{2}}}$. $\epsilon_{\text{\tiny \textbf{1}}}$ and $\epsilon_{\text{\tiny \textbf{2}}}$ are random errors due to limited resolution in generating time delay  $\delta$.
From Fig.~\ref{fig:SDSTWRList} RTT measurement at node \textbf{1} is \begin{equation}
 y_{\text{\tiny \textbf{1}}}=2 \frac{\rho_{\text{\tiny \textbf{12}}}}{c}(1+\vartheta{\text{\tiny \textbf{1}}}) +\Delta_{\text{\tiny \textbf{2}}}  (1+\vartheta{\text{\tiny \textbf{1}}}-\vartheta{\text{\tiny \textbf{2}}})+ \eta,
 \label{eq:ALICE}
 \end{equation}
    where $\eta$ is AWGN with variance $\sigma^2=2\sigma_j^2+2\sigma_c^2$. $\sigma_j^2$ is the variance of the AWGN jitter noise of the clocks $\mathcal{C}_{\text{\tiny 1}}$ and $\mathcal{C}_{\text{\tiny 2}}$ and  $\sigma_c^2$ is  the variance of the AWGN channel noise \cite{DwivediEtAl2015_jointsyncrange}. The term $2\rho_{\text{\tiny \textbf{12}}}\vartheta{\text{\tiny \textbf{1}}}$ is the error in measuring propagation delay by $\mathcal{C}_{\text{\tiny \textbf{1}}}$ due to its clock skew $\vartheta{\text{\tiny \textbf{1}}}$. The delay $\Delta_{\text{\tiny \textbf{2}}}$ generated by $\mathcal{C}_{\text{\tiny \textbf{2}}}$ and measured by node \textbf{1} as a part of RTT measurement is  $\Delta_{\text{\tiny 2}}(1+\vartheta_{\text{\tiny 1}}-\vartheta_{\text{\tiny 2}})$. RTT measurements by node \textbf{2} can be written as above. 

We see from Fig.~\ref{fig:SDSTWRList} that the node \textbf{3} listens to the exchanges between node \textbf{1} and node \textbf{2}. Node \textbf{3}   makes  measurements $y_{\text{\tiny \textbf{3}}}^{\text{\tiny \textbf{12}}}$ and $y_{\text{\tiny \textbf{3}}}^{\text{\tiny \textbf{21}}}$.  
Measurement models at node \textbf{3} are,
\begin{eqnarray}
\!y_{\text{\tiny 3}}^{\text{\tiny 12}} \!=\! \frac{1}{c}\left(  \rho_{\text{\tiny 12}}+\rho_{\text{\tiny 23}} -\rho_{\text{\tiny 13}}\right) (1+ \vartheta{\text{\tiny 3}}) +\Delta_{\text{\tiny 2}}(1+\vartheta{\text{\tiny 3}}-\vartheta{\text{\tiny 2}})  + \eta^{\text{\tiny 12}}\!\!\!\!, \label{eq:yea} \\
\!y_{\text{\tiny 3}}^{\text{\tiny 21}}\!=\!\frac{1}{c}\left(  \rho_{\text{\tiny 12}} + \rho_{\text{\tiny 13}} -\rho_{\text{\tiny 23}} \right) (1+ \vartheta{\text{\tiny 3}})+\Delta_{\text{\tiny 1}}(1+\vartheta{\text{\tiny 3}}-\vartheta{\text{\tiny 1}}) + \eta^{\text{\tiny 21}}\!\!\!\!.\label{eq:yeb}
\end{eqnarray}
$\eta^{\text{\tiny 12}}$ and $\eta^{\text{\tiny 21}}$ are AWGN noise sources in measurements with variance $\sigma^2=2 \sigma_j^2+ 2\sigma^2_c$. 

The sum of above two measurements can be written as
\begin{eqnarray}
y_{\text{\tiny 3}}^{\text{\tiny 12}}+y_{\text{\tiny 3}}^{\text{\tiny 21}}  =  2  \frac{\rho_{\text{\tiny 12}}}{c}(1+\vartheta{\text{\tiny 3}}) + \delta(2 + 2 \vartheta{\text{\tiny 3}} - \vartheta{\text{\tiny 1}} - \vartheta{\text{\tiny 2}}) +\nonumber\\
 \epsilon_{\text{\tiny 1}}(1+\vartheta{\text{\tiny 3}}-\vartheta{\text{\tiny 1}}) + \epsilon_{\text{\tiny 2}}(1+\vartheta{\text{\tiny 3}}-\vartheta{\text{\tiny 2}}) + \eta^{\text{\tiny 12}} + \eta^{\text{\tiny 21}} 
\label{eqlisten}
\end{eqnarray}

\begin{figure}[h!]
        \centering
        \includegraphics[width=0.7\columnwidth]{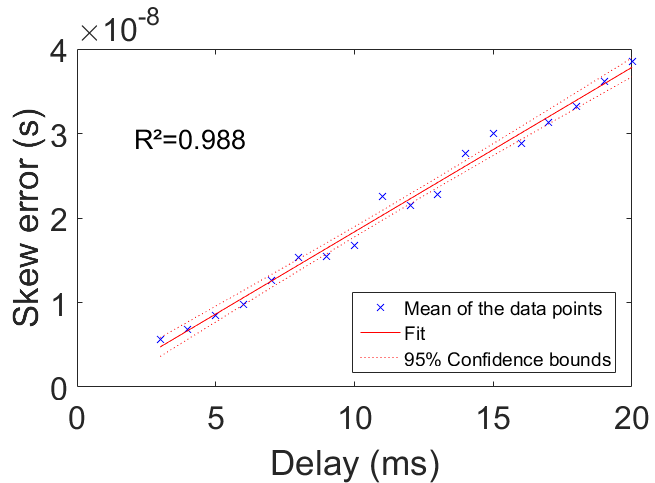}
        \caption{Experimental behaviour of the skew error in TWR for a 1 m distance with  $\delta$ varying between 3 and 20 ms.}
        \label{fig:Delay}
    \end{figure}

 From (\ref{eqlisten}), the listener node \textbf{3} can estimate $\rho_{\text{\tiny 12}}$ as long as effects due to clock error parameters $(\vartheta{\text{\tiny 1}}, \vartheta{\text{\tiny 2}}, \vartheta{\text{\tiny 3}}, \epsilon_{\text{\tiny 1}}, \epsilon_{\text{\tiny 2}})$ remain sufficiently small. As can be seen from (\ref{eqlisten}), the biggest error term is the skew error, $\delta(2\vartheta{\text{\tiny 3}}-\vartheta{\text{\tiny 1}}-\vartheta{\text{\tiny 2}})$ which increases linearly with $\delta$ as is evident from experimental observation in  Fig.~\ref{fig:Delay}.

\subsection{Measurement model for schedule based localization}
In this sub-section, we extend the measurement model for self localization presented in \cite{6490337}. We introduce clock error terms discussed above as they show up in practical implementation of schedule based self-localization setup. Consider a network of $N$ anchor nodes with a self localizing node '\textbf{L}' with coordinates $\mathbf{x}=[(x_1, y_1), (x_2, y_2) .. (x_N, y_N) , (x_L, y_L)]^{\text{\tiny T}} $. The vector of ranges, $\boldsymbol{\rho}_{\text{\tiny L}} =[\rho_{\text{\tiny 12}},..., \rho_{\text{\tiny N-1,N}},\rho_{\text{\tiny L1}}, ... ,\rho_{\text{\tiny LN}}]^{\text{\tiny T}}$.The relative skew between $\mathbf{L}$ and a node $i$, $\vartheta_\text{\tiny Li} = \vartheta_\text{\tiny L} - \vartheta_\text{\tiny i}$.

 For a specific schedule of $M $ measurements, we can compute the matrix $\mathbf{S}$,
 of size $M \times \frac{N(N-1)}{2}+N$  which maps the time of flights to the measurements. A single TDOA measurement at the listener node \textbf{L} between successive reception of signals from node $i$ and node $j$ follows (\ref{eq:yea}).  The resulting row in $\mathbf{S}$ being \begin{equation}\mathbf{S}_{\text{\tiny L}}^{\text{\tiny ij}}=[ 0, ..., \underbrace{1}_{\mathclap{ \rho_{ij}}},0, ...,\underbrace{-1}_{\mathclap{ \rho_{iL}}},0, ...,\underbrace{1}_{\mathclap{ \rho_{jL}}},0,...,0].
\end{equation}
Then based on (\ref{eq:yea}) considering that all the noise sources $\eta$ are AWGN, independent and identically distributed the measurement vector at \textbf{L} is
\begin{equation}\label{eq:vectTrue}
\mathbf{y_{\text{\tiny L}}} =\frac{1}{c}\:\mathbf{S} \: \boldsymbol{\rho}_{\text{\tiny L}} +\mathbf{D} +\frac{1}{c}\: \vartheta{\text{\tiny L}}\: \mathbf{S}  \boldsymbol{\rho}_{\text{\tiny L}}  + \mathbf{R}\:\mathbf{D} + (\mathbf{I}+\mathbf{R})\:\boldsymbol{\epsilon}+\boldsymbol{\eta}\qquad \in \mathbb{R}^M .
\end{equation}
Where, $\mathbf{RD}=\text{Diag}(\vartheta_\text{\tiny Li}) \mathbf{1} \delta $ is the vector of skew errors ordered as per the schedule, $\mathbf{I}$ the identity matrix of the measurement space, $\boldsymbol{\eta}$ the noise vector with covariance matrix $\textbf{Q}=\text{Diag}(\sigma^2)$ and $\boldsymbol{\epsilon}$ is the vector of sending delay errors $\epsilon$.

Consider the schedule between 3 anchors used for positioning in \cite{6340491}: $[1, \: 2,\: 3,\: 2,\: 1,\: 3] $, where $M = 5$, $\boldsymbol{\rho}_\text{\tiny L}=[\rho_\text{\tiny 12},\rho_\text{\tiny 13},\rho_\text{\tiny 23},\rho_\text{\tiny L1},\rho_\text{\tiny L2},\rho_\text{\tiny L3}]^\text{\tiny T} $, then $ \mathbf{S}\in \mathbb{R}^{5\times 6}$, $\mathbf{D}=\delta\mathbf{1} $ ,$\mathbf{RD}=\delta[\vartheta_\text{\tiny L2},\vartheta_\text{\tiny L3},\vartheta_\text{\tiny L2},\vartheta_\text{\tiny L1},\vartheta_\text{\tiny L3}]^{\text{\tiny T}}$, and $\boldsymbol{\epsilon}=[\epsilon_1,\epsilon_2,\epsilon_3,\epsilon_4,\epsilon_5]^{\text{\tiny T}}.$

\section{Clock Error Mitigation}
Clock errors in in the measurement can be reduced by two ways. The error $\boldsymbol{\epsilon}$ can be minimised by transmitting actual delays $\boldsymbol{\Delta}$. In addition, relative skews of node \textbf{L} and anchor nodes, $\mathbf{R}$, can be estimated from scheduled measurements.
\subsection{Error mitigation by transmission of the actual delay}
\label{sec:payload}
\begin{figure}
   \centering
\begin{subfigure}[b]{0.49\columnwidth}
              \includegraphics[width=0.9\columnwidth]{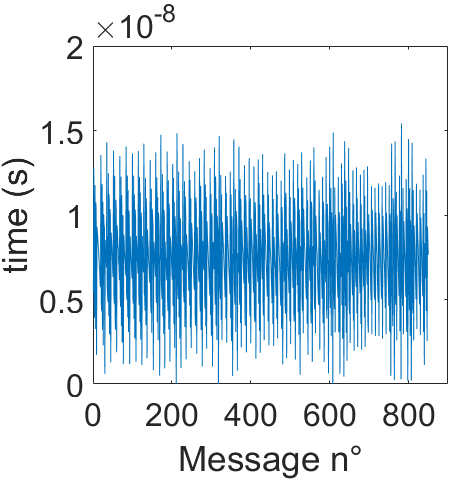}
                \caption{Measurements  without dynamic retrieval of the delay}
                \label{fig:noretrieval}
        \end{subfigure}%
        \hfill
       \begin{subfigure}[b]{0.49\linewidth}
            \includegraphics[width=0.9\columnwidth]{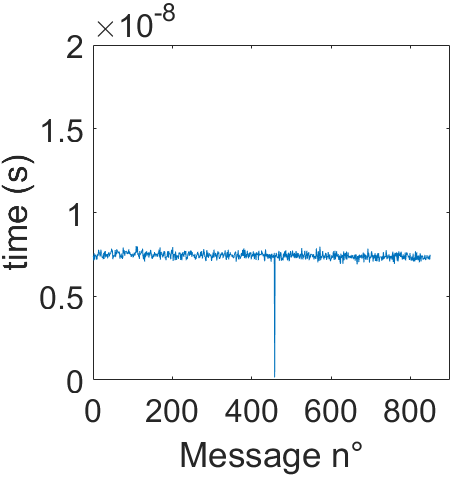}
                \caption{Measurements with dynamic retrieval of the delay}
                \label{fig:withretrieval}
        \end{subfigure}%
        \caption{Experimental study of the impact of dynamic retrieval of the observed delay}\label{fig:DelayImpact}
    \end{figure}

To mitigate the clock error $\boldsymbol{\epsilon}$, the transmitting node in the schedule transmits $\boldsymbol{\Delta}$ in the subsequent transmission. The $\Delta$ is transmitted as payload in the broadcast message by the transmitting node.  This leads to an addition of a few bytes in the total message frame, but provides the listener node the exact delay generated at the sender. The direct retrieval of $\Delta$ reduces the effect of random error $\epsilon$. The influence of this dynamic retrieval can be seen in Fig.~\ref{fig:DelayImpact}, in the presented measurement set, this retrieval leads to a reduction of the standard deviation  from $3.3235$\! ns  to $0.30666$\! ns.


\begin{figure}[ht]
       \centering
       \begin{subfigure}[b]{0.49\columnwidth}
              \includegraphics[width=\columnwidth]{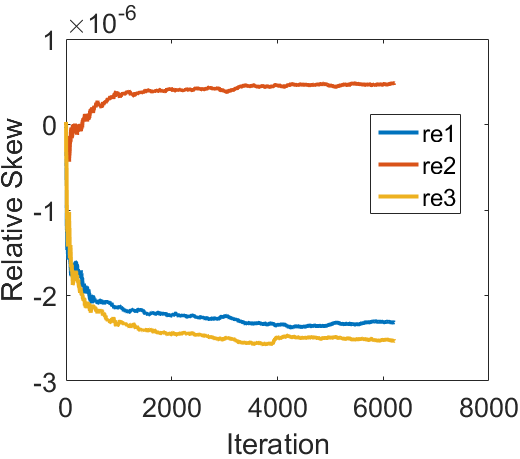}
                \caption{RLS applied on the measurements without dynamic retrieval of the observed delay}
                \label{fig:noretrieval}
        \end{subfigure}%
        \hfill
       \begin{subfigure}[b]{0.49\linewidth}
            \includegraphics[width=\columnwidth]{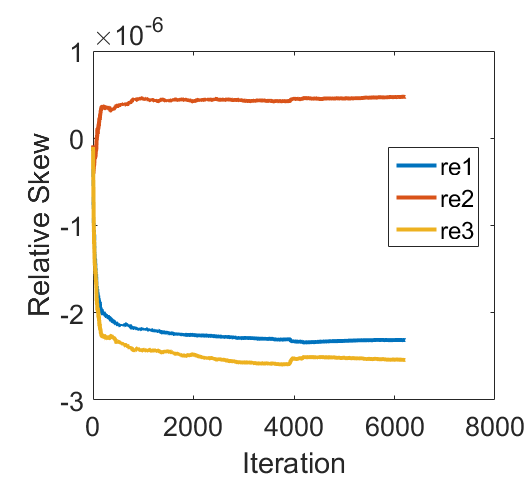}
                \caption{RLS applied on the measurements with dynamic retrieval of the observed delay}
                \label{fig:withretrieval}
        \end{subfigure}%
        \caption{RLS estimation of the relative clock skew on experimental data}
        \label{fig:RLSSkew}
    \end{figure}

\subsection{Inline calibration in Scheduled based self-localization}
\label{sec:RLS}

As shown in the TWR case, the biggest error term in a scheduled based self localisation measurement is the skew error $\mathbf{R}\mathbf{D}$ that appears as a bias. Considering a minimal length valid schedule, $\mathbf{S} $ is then full rank and $\mathbf{S} \in \mathbb{R}^{N(N-1)/2+N-1 \times(N-1)/2+N}$ \cite{Zachariah2014}, and by construction  $\mathbf{u} = [0_{\text{\tiny 12}},\cdots, 0_{\text{\tiny N-1N}},1_{\text{\tiny L1}},\cdots, 1_{\text{\tiny LN}}]^{\text{\tiny T}} \in \text{ker}\: \mathbf{S}$, hence as $ \text{dim}(\text{ker} \mathbf{S})=1 $, $\text{ker}\:\mathbf{S} = \text{span}(\mathbf{u})$
 leading all the distances between the anchors to be identifiable in our measurements. Let $\mathbf{S}^+$ be the Moore-Penrose pseudo inverse of $\mathbf{S}$ and $\boldsymbol{\Pi}=\left(\begin{array}{cc}
    \mathbf{I}_{\frac{N(N-1)}{2}} & \mathbf{0} \\
    \mathbf{0} & \mathbf{0}
\end{array}\right)$,we have $ \boldsymbol{\Pi} \mathbf{S}^+ \mathbf{S} = \boldsymbol{\Pi}  $, then under our previous assumptions
\begin{equation}\label{eq:used}
\boldsymbol{\Pi}\:\mathbf{S}^+\mathbf{y_{\text{\tiny L}}} \simeq \frac{1}{c}\left(\begin{array}{c}
   \rho_{\text{\tiny 12}}\\\cdots\\ \rho_{\text{\tiny N-1N}}  \\ \mathbf{0}
    
\end{array}\right)+\boldsymbol{\Pi}\:\mathbf{S}^+\left(\mathbf{I}+\mathbf{R}\right) \mathbf{D}.
\end{equation}
As $\text{range}(\boldsymbol{\Pi})$ is isomorphic to $\mathbb{R}^{\frac{N(N-1)}{2}}$ we can then resolve in $\mathbb{R}^{\frac{N(N-1)}{2}}$ the N unknown $\vartheta{\text{\tiny L}}-\vartheta{\text{\tiny Sender}}$ of $\mathbf{R}$, as long as the distance between the anchors and $\mathbf{D}$ are known. As we are considering a schedule composed of anchors using known delays these conditions are fulfilled and inline calibration is possible. 

We propose a recursive estimator which updates skew error estimates with every new measurement. The framework is based on recursive least squares (RLS) estimation \cite{Kay1993_ssp}.

 Let $\boldsymbol{\theta}=\left[\vartheta{\text{\tiny L1}}, \cdots, \vartheta_{\text{\tiny LN}}\right]^{\text{\tiny T}}$ be the vector of the relative skews, then (\ref{eq:used}) can be rewritten 
\begin{equation}
\boldsymbol{\Pi}\mathbf{S}^+\mathbf{R} \mathbf{D}=\boldsymbol{\Pi}\left(\mathbf{S}^+\cdot\left(\mathbf{y}_{\text{\tiny L}}-\textbf{D}\right)- \boldsymbol{\rho}_{\text{\tiny L}}\right)=\textbf{G}^T \: \boldsymbol{\theta},
\end{equation}
where $\textbf{G}$ is a matrix that maps the relative skews $\boldsymbol{\theta}$ into the vector $ \boldsymbol{\Pi}\mathbf{S}^+ \mathbf{R} \mathbf{D}$.
Both $\textbf{G} $, $\textbf{S}^+$, $\mathbf{D} $ and $\boldsymbol{\rho}_{\text{\tiny L}}$ can be precomputed as they only depend on anchor position and the schedule.
By ignoring the last N dimensions, we can estimate $\boldsymbol{\theta}$ by RLS(with a unit forgetting factor), see Algorithm 1.
    \begin{algorithm}
\begin{algorithmic}
\REQUIRE previous estimate $\widehat{\boldsymbol{\theta}}_n$, previous inverse covariance estimation $\textbf{P}^{-1}_n$, system matrix $\textbf{G}$, measurement update $\mathbf{d}_n$

\STATE  $\mathbf{y}_{n+1} = \textbf{G}^\text{\tiny T}\cdot\widehat{\boldsymbol{\theta}}_n$
\STATE  $\textbf{K}=\textbf{P}^{-1}_n \textbf{G}^*\left[\textbf{I}+\textbf{G}^*\textbf{P}^{-1}_n\textbf{G}\right]^{-1}$
\STATE $\widehat{\boldsymbol{\theta}}_{n+1}= \widehat{\boldsymbol{\theta}}_n + \textbf{K}\left(\mathbf{d}_n-\mathbf{y}_{n+1}\right) $
\STATE $\textbf{P}^{-1}_{n+1}=\textbf{P}^{-1}_n+\textbf{K}\textbf{G}^\text{\tiny T}\textbf{P}^{-1}_n$

\ENSURE $\textbf{P}^{-1}_{n+1}$,$\widehat{\boldsymbol{\theta}}_{n+1}$
\end{algorithmic}
\label{RLSE}
\caption{Recursive Least Square estimator}
\end{algorithm}

$\textbf{P}^{-1}_{n+1}$ and $\textbf{K}$ can be computed offline
, resulting in a quadratic cost in the number of anchors for each iteration.

Fig.~\ref{fig:RLSSkew} shows the evolution of the estimated relative skews for two types of data, the first one using raw data, the second one using the same data with variance reduced by using information on the delay error. It is clear that the variance reduction induces a faster convergence of the RLS. Both estimations converging towards the same values.

    Clock error mitigation from the measurements reduces the measurement model in (\ref{eq:vectTrue}) to the    approximate model \begin{equation}
    \mathbf{y}_{\text{\tiny L}}=\frac{1}{c}\mathbf{S}\: \boldsymbol{\rho}_{\text{\tiny L}} + \mathbf{D}+\mathbf{w},
    \label{approx}
    \end{equation}
    where $ \mathbf{w}  \sim \mathcal{N}(0,\sigma^2 \mathbf{I}) $.
     To perform localization we aim to maximize $p(\mathbf{x},\sigma^2|\mathbf{y}_{\text{\tiny L}})$, this is equivalent to minimize \cite{6490337}, \begin{equation}
        \mathbf{V}(\mathbf{x})= \frac{1}{2} \text{ln} \parallel \mathbf{y}_{\text{\tiny L}} -\frac{1}{c} \mathbf{S}\:\mathbf{g}(\mathbf{x}) -\mathbf{D}\parallel^2 + \frac{\beta}{2}\parallel \boldsymbol{\mu} - \mathbf{x} \parallel^2_{\mathbf{Pr}^{-1}}.
    \end{equation}
    
    Where $\mathbf{g}$ is the function that maps the positions onto the distances, $\beta=1/(M+2) $. The vector $\boldsymbol{\mu}=[\mathbf{\mu}_1,\cdots,\mathbf{\mu}_N,\mathbf{\mu}_{\text{\tiny L}}] $  is mean of the prior on node positions. 
    $\mathbf{Pr}^{-1} $
    is the inverse of covariance matrix on node positions. The position is estimated through the iterative MAP estimator proposed in \cite{6490337}. The Hybrid Cramér Rao Bound (HCRB) can then be computed as $\mathbf{C}  	\succeq \mathbf{J}^{-1}$ where $\mathbf{J}=\mathbf{J}^D+\mathbf{J}^P$ \cite{vanTrees2004_detection}. $\mathbf{J}^D$ being the expected Fisher information of the parameters \cite{6490337}.
\begin{equation}
\text{J}^D_{i,j}=\frac{1}{c^2 \sigma^2} \frac{\partial g^\text{T}}{\partial x_i}\mathbf{S}^\text{T}\mathbf{S}\frac{\partial g}{\partial x_j}+ \frac{M}{2\sigma^4} \frac{\partial \sigma}{\partial x_i}\frac{\partial \sigma}{\partial x_j}.
\end{equation}
$\mathbf{J}^P$ is the inverse matrix of prior covariances $\mathbf{Pr}^{-1}$.
 \section{Experimental study on self localization}   

\begin{figure}[ht]
        \centering
        \includegraphics[width=0.8\columnwidth]{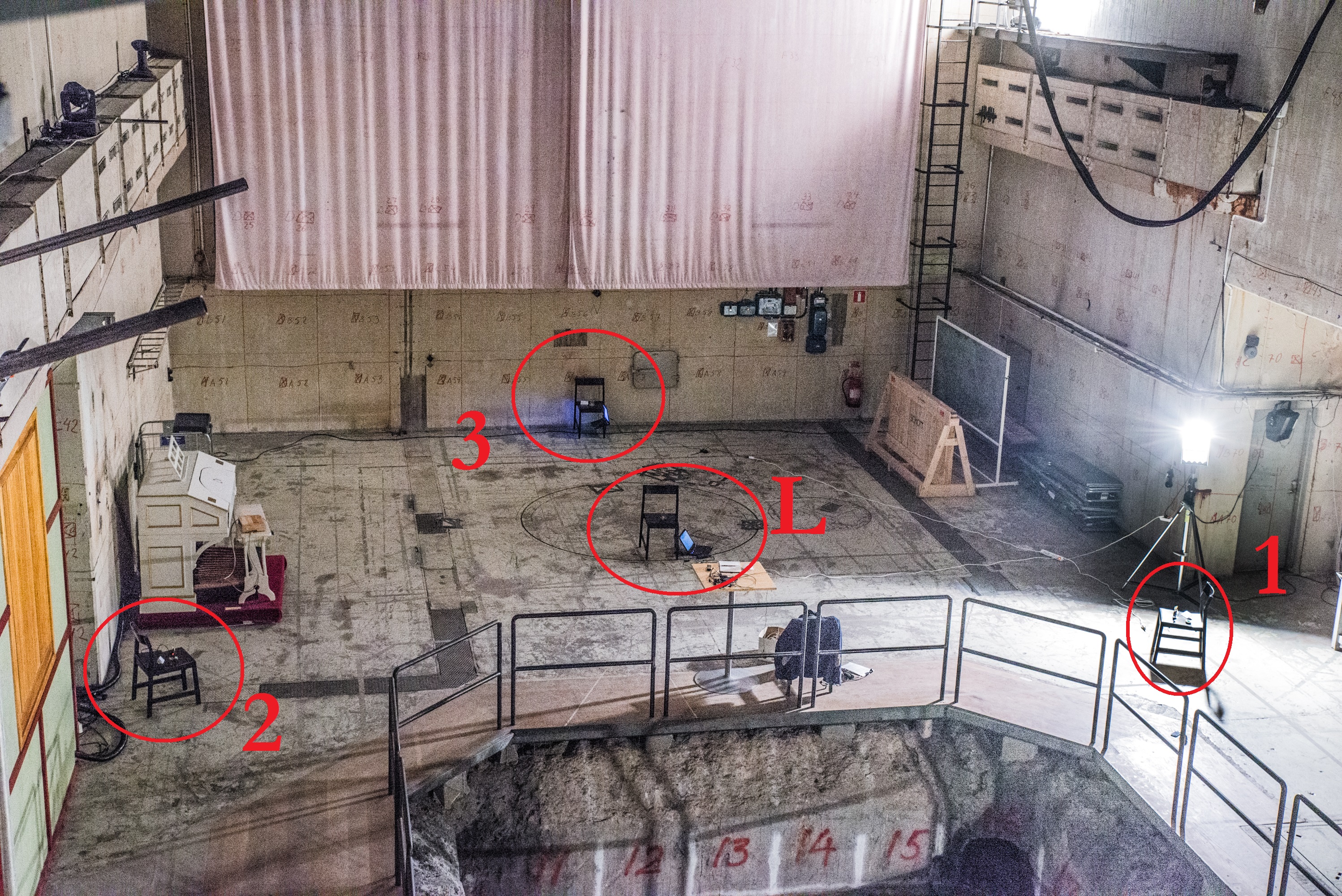}
        \caption{Experimental setup in Reactorhallen R1, where 1, 2 and 3 are anchors and L the listener node}
        \label{fig:setup}
    \end{figure}

In order to test the quality of the proposed model and calibration method, we set up an indoor positioning experiment (See Fig.~\ref{fig:setup}). In this experiment we have 3 anchors separated by approximately 10 meters participating in a schedule with a 3 ms delay ($\delta$) and a self localizing listener node using the MAP estimator. 
In the presented scenario, the 4 nodes are DW1000 radios commercialised by Decawave Ltd. The messages used follow IEEE standard 802.15.04-2011 with 18 coded octet and a header 1024 symbols, a pulse repetition frequency (PRF) of 16 MHz. The channel used has a  3993.6 MHz centre frequency and a bandwidth of 1331.2 MHz.
The data have been gathered in line of sight (LOS) conditions. For outlier rejection, based on that the distances should not exceed 20 m in our scenario, a whole schedule is discarded as soon as one measurement of $\left|y_{\text{\tiny L}}-\Delta \right|$ 
exceeds 100 ns.
The schedule used is $[1, 2, 3 ,2, 1, 3]$. Our specific implementation of scheduled based positioning allows one more measurement as the last node also waits $\delta$ after reception of the last message. This measurement leads the estimation to be based on a schedule of $[1, 2, 3 ,2, 1, 3, 1]$.

\begin{figure}[ht]
        \centering
        \includegraphics[width=0.8\columnwidth]{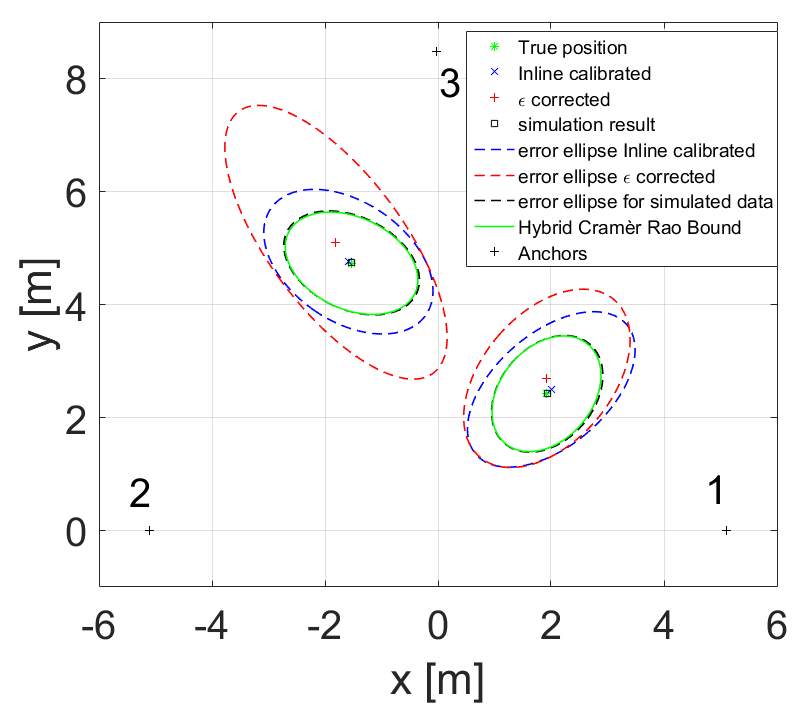}
        \caption{Experimental results of passive localization for 2 different positions $\{1.92,2.42\}$ and $\{-1.53,4.73\}$, the simulated data and hybrid Cramér Rao Bound with zero mean Gaussian noise of standard deviation $\sigma= 3$ ns.}
        \label{fig:resultposit}
    \end{figure}

Fig.~\ref{fig:resultposit} presents the results for self-localization in the Fig.~\ref{fig:setup} setup. The measurements are collected at two different locations of node \textbf{L}. The collected measurements are further used for self-localization using the MAP estimator discussed previously. Fig.~\ref{fig:resultposit} shows the following error ellipses with $99\%$ confidence interval
\begin{enumerate}
\item Error ellipse for HCRB. As seen its the smallest ellipse with lower bound on position error. 
\item Error ellipse simulation setup. The mean of the error ellipse is within $2$ centimeters of the true node position.
\item Error ellipse for position estimation by MAP estimator with the measurements obtained by error correction after direct retrieval of real delays.
Mean of the estimated position lies away from true positions by a distance of $27$ and $46$ centimeters.
\item Error ellipse for position estimation by MAP estimator with the measurements obtained by inline calibrated measurements where clock errors have been minimized. Mean of the estimator lies away from true position by $9$ and $6$ centimeters.
\end{enumerate}
 As the measurements without any clock error mitigation has high variance and bias, the estimator does not converge to something realistic, hence results are not shown for this case.

   The results after removing clock errors from the measurements  are close to simulation results. However, many other unaccounted and practical issues such as presence of outliers in the original measurements and some discrepancy between simulation parameters and corresponding values in experimental setup results in slight difference in performances.

\section{Conclusion}

We have proposed a practical way to implement schedule based self-localization.
We have arrived at a measurement model accounting for real world errors in making the self localization work. The clock errors dominate the non-idealities in the measurements, hence efforts were put in to suggest different ways of mitigating clock errors. The primary suggestions for mitigating clock errors while experimenting with Decawave devices are transmission of delays along with message by the transmitter, and estimating relative clock skews of nodes in the context of schedule based self localization. The results validate suggested methods while showing mean of estimate within $10$ centimeters of true position.

\bibliographystyle{IEEEbib}
\bibliography{ref}

\end{document}